\documentclass{aa}
\usepackage{graphics}
\begin{document}
\thesaurus{06     
              (03.11.1;  
               16.06.1;  
               19.06.1;  
               19.37.1;  
               19.53.1;  
               19.63.1)} 
\title{Search for Non-Triggered Gamma Ray Bursts in the BATSE 
Continuous Records: Preliminary Results}
\author{B.Stern\inst{1,} \inst{4}, Ya.Tikhomirova \inst{2}, M.Stepanov, \inst{3} 
D.Kompaneets \inst{2}, A.Berezhnoy \inst{3}, R.Svensson \inst{4}} 

\institute {${}^1$Institute for Nuclear Research, Russian Academy of Sciences
Moscow 117312, Russia, stern@al20.inr.troitsk.ru\\
${}^2$Astro Space Center of Lebedev Physical Institute,
Moscow, Profsoyuznaya 84/32, 117810, Russia\\
${}^3$ Skobeltsyn Institute of Nuclear Physics, 
       Moscow State University, 
       Moscow 119899, Russia\\
${}^4$Stockholm Observatory, S-133 36 Saltsj\"obaden, Sweden}
\date{Received ..........; accepted .......}
   
\authorrunning {Stern {\it et al}}
\titlerunning{1200 Non-triggered GRB}

\maketitle

\begin{abstract}

We present preliminary results of an off-line search for non-triggered
gamma-ray bursts (GRBs) in the
BATSE daily records for about 5.7 years of observations.
We found more GRB-like events than the yield of the similar search of 
Kommers et al. (1998)
and extended the Log N - log P distribution down to $\sim$ 0.1 ph cm$^{-2}$ 
s$^{-1}$. The indication of a turnover of the log N - log P at a small P
is not confirmed: the distribution is straight at 1.5 decades
with the power law index -.6
and cannot be fitted with a standard candle cosmological model.

\end{abstract}

\section{Introduction}
 Many gamma-ray bursts which were too weak to cause the BATSE to trigger
or were being missed due to other reasons (data readouts etc.)
can be 
confidently identified 
in the BATSE daily records which cover the full period of the CGRO operation. 

 The search for non-triggered bursts can be of crucial importance in the 
following respects:

 - Extension of the log N - log P distribution which is necessary to make  
conclusive cosmological fits.

 - To reveal different GRB subpopulations, should they exist. In particular,
the weak end of log N - log P can reveal subpopulations with Eucledian
brightness distributions or put strong constraints on them.

 - To refine angular distributions and constraints on anisotropy
associated with the Galaxy and M31. This in turn can put much stronger 
constraints on the Galactic halo scenario.

 - To increase the probability to observe the gravitational lensing 
effect in GRBs by more than a factor of 2.

 The systematic search for non-triggered GRBs was started by Kommers et al.
(1997). Recently, Kommers et al.(1998) completed the data scan for
6 years.

  Despite that we started our search (in November, 1997) for non-triggered
bursts much later than Kommers et al.(1997), our work still have several
strong motivations:

 - Any scientific work subjected to difficult selective biases becomes much 
more reliable when performed independently by different groups.

 - We started our scan with some important advances. Firstly, we used a
more selective off-line trigger code. Secondly, and most importantly, is the
method of measurement of the efficiency of the GRB search using artificial test 
bursts that we employed. 

 More than a half of the human power was spent on the finding and 
the fitting of artificial test bursts for the sake of testing the 
reliability of log N - log P  distribution of real bursts.


\section{Data scan}

 We use 1024 ms time resolution BATSE data (DISCLA) from the ftp 
archive of the Goddard Space Flight Center at 
ftp://cossc.gsfc.nasa.gov/compton/data/batse/daily/.

The procedure of data reduction contains the following steps: 

 Step 1. {\bf Conversion} of the original BATSE records adding to them 
artificial test bursts prepared from real rescaled bursts taken from
the BATSE database.  

The number of test bursts in the sample is 500. All are longer than 1 s.

 Each test burst was made by randomly sampling one of the 500 bursts and 
rescaling its amplitude
 to the randomly sampled peak count rate with a proper Poisson noise 
(the lower limit being 160 counts
/s/2500cm$^2$ , which approximately corresponds to $\sim$ 0.1 ph 
cm$^{-2}$ s$^{-1}$)
 
 Test bursts were added to the data at a random time 
with an average time interval 25000 s
(i.e., the  number of test bursts 
exceeds the number of real bursts).

\vspace{0.5cm}

 Step 2. {\bf Data scan}

 We performed automatic check for the trigger conditions (see sec.3).

 Each trigger was followed by a human decision whether the trigger is a GRB
candidate. The decision was made using:

- residual $\chi^2$, hardness ratio and other quantities;

- count rate curves in different detectors and energy channels;

- a$\chi^2$ map of the sky for the event (residual $\chi^2$ after fit from a given 
direction) with projected Sun, Cyg X-1 and Earth horizon.

All candidates were recorded as a fragments of daily records
saving all original information.

 The person performing the scan was unaware whether it was a real or a test 
burst.    

\vspace{0.5cm}

 Step 3. {\bf Event classification.}
 
 The candidate events were discarded or classified as non-GRBs using the 
following criteria:

 - A low statistical significance (using a $\chi^2$ map of the event over the 
sky: $\chi^2$ should exceed $4 \sigma$ over its minimum value in a 
hemisphere opposite to the best fit direction.)

 - A bad directional fit: strong signals in all detectors, a bad $\chi^2$ map
(many local minima),
a spike in a 
single detector (luminiscence from heavy nuclei).

 - Soft, appeared during high ionosperic activity, and is close or below
to the
horizon (ionospheric events).

 - Soft and close to the Sun (solar flare) or 
consistent with Cyg X-1 or another known x-ray sources in the corresponding
location and is of the  same range of intensity and hardness.

\vspace{0.5cm}

 Step 4. {\bf Separation of tests bursts} using the protocols 
generated on step 1. 

\section{Off-line trigger}

Background estimation for the trigger was done by linear fit of count rate 
over preceeding 40 seconds. (The backround estimate for BATSE trigger is 
the average over preceeding 17 seconds. Kommers et al.(1997)
 use linear fit 
in interval dependind on triggering time scale.)

Trigger criteria were the following (all criteria should be satisfied 
simultaneously):

1. The first criterium was traditional: a significant count rate excess 
over background: brightest detector - $4\sigma$ excess,
  second brightest - $2.5\sigma$.
The excess was checked in time intervals (triggering timescales) 
1 bin (1.024s), 2 bins, 4 bins, and 8 bins. Count rates in energy channels
\#2 and \#3 (50 - 300 keV) were used for triggering. 
(Kommers et al. 1997  used similar thresholds in
the same time intervals.
BATSE trigger was set up to 5.5$ \sigma$ and 5.5 $\sigma$ correspondingly, 
sometimes higher.
)

2. The second criterium was a test on sufficient time variability
using $\chi^2$ threshold over intervals around triggering time.
 After fitting the signal summed over all triggered detectors by a 
straight line in the intervals 
-16 s $< T <$ 16 s and -16 s $< T <$ 32 s, the residual
 $\chi^2$ in one of intervals should 
exceed 2.5 per degree of freedom.
 The threshold value was chosen using a sample of weak non-triggered burst 
found before applying this criterium: all of them passed this threshold.

 The criterium is efficient against false triggers on smooth background
variations and occultation steps. This criterium was not applied by 
Kommers et al.(1997). 

3. The third criterium was based on Cyg X-1 subtraction.
 The detector counts are fitted using a signal incident from Cyg X-1
direction (no real detector response matrix is used in this step -- just  
cos$(\theta)$ factor). Then the residual count rate pattern is checked
for sufficient variability with $\chi^2$ threshold as in the previous step.
If $\chi^2$ exceeds threshold for one of a few test time intervals,
the trigger is accepted and the scanning code stops for human 
interactive operation.  

 Criteria (2) and (3), which were not used before, turned out to be 
very 
efficient against false triggers reducing their number from hundreds 
to several per day for quite Cyg X-1 or to few tens for loud Cyg X-1. 
These criteria can reject 
some real bursts but their number should be small: a very smooth
profile is not tipical for GRBs and Cyg X-1 subtraction cuts out 
less than 5\% of the sky. On the other hand
such criteria improve the efficiency of the human visual
stage 
of the work. This is a possible reason why we have found more non-triggered 
events than Kommers et al., (1998)

\section{Preliminary results}

 We scanned 2068 days of BATSE daily records and found
1243 non-triggered events which can be classified as classic GRBs.
 (Kommers et al.,(1998) found 837 non-triggered GRBs 
per 2200 days). We found 1374 bursts which were triggered by 
BATSE 
(Kommers et al.,(1998) detected 1393 BATSE triggered events), and
missed near 350 BATSE triggers: some of them are in data gaps, 
some are too
short to be detected with 1 s time resolution.

Because many short bursts are lost in 1 s resolution daily
BATSE data our scan yielded mostly long ($>$1 s) events.

 We also found 3780 test bursts out of about 6800 added to 
 the data. 

 The comparison with catalog of Kommers et al. (1997)for
 one year of observations
showed that we found 90\% of their events and 
approximately the same amount missing in their catalog. 
Events that we did found and Kommers et al.(1997) did not
are not necessary the weakest bursts.
Later, Kommers et al. (1998) increased their efficiency, so our final 
statisticsis about 50\% larger.

 The peak flux distribution of events found in the scan is presented in Fig.1.
\begin{figure}
  \resizebox{\hsize}{!}{\includegraphics{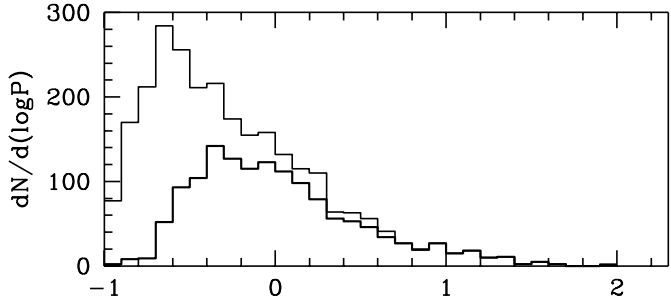}}
\caption{ Differential peak flux distribution of detected GRBs.
  Thick histogram -- the distribution of BATSE triggers detected
in the scan (1374). Thin histogram -- all bursts detected in the scan
 (2617). Both distributions are given before correction for the 
efficiency is applied.}
\end{figure}
Note that BATSE trigger missed some events much above threshold due to
readout dead time. The efficiency measured by test bursts is shown in Fig.2.

\begin{figure}
  \resizebox{\hsize}{!}{\includegraphics{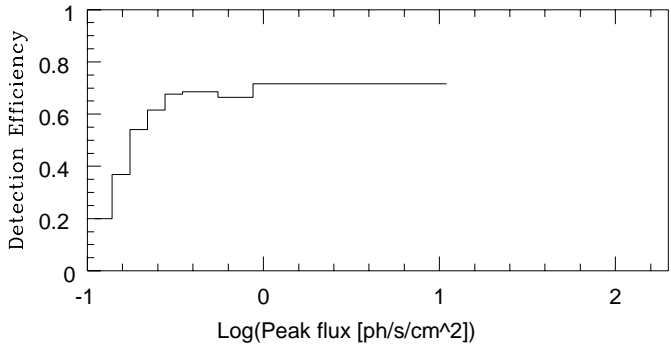}}
\caption{ The efficiency of the off-line burst detection defined as a fraction of 
test bursts detected in our scan versus expected count rate. 
 The statistic of strong test bursts is poor
as the main fraction of test bursts were sampled at low brightness
in order to develop declining part of the efficiency curve.
With this reason we averaged the efficiency over a wide interval at the
large peak count rates. The asymthotic efficiency (0.72) refers to the
total number of events appearing above the Earth horizon. }
\end{figure}

Data gaps and periods of a high ionospheric background are taken into
account so 
the efficiency is normalized to whole elapsed time of CGRO operation.

 Angular distributions of new events show no excess towards the Sun, the excess 
$\sim 20$ events in the direction of Cyg X-1 and reasonable 
distribution in Geocentric
coordinates (a smoothed step at Earth horizon and isotropy above it which 
indicates that a possible contamination of our sample with 
ionospheric events is small).
The equatorial angular distribution is consistent with the sky coverage
function given by Meegan et al (1998).

The hardness - peak flux scattering plot shown in Fig. 3 demonstrates that
new weak bursts give a direct continuation of the distribution of stronger GRBs.
All possible backgrounds events are softer on average.


 \begin{figure}
  \resizebox{\hsize}{!}{\includegraphics{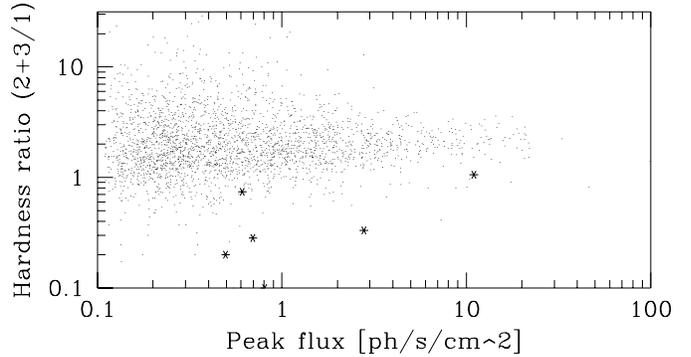}}
\caption{Hardness - peak flux plot for all detected GRBs.
Note that weak non-triggered bursts give a
direct continuation of the distribution of stronger GRBs.
The well known hardness-brightness correlation  is clearly
visible. The high hardness ratios for weak bursts often  are statistical
fluctuations while low hardness events are real. There are no solar
flares in this plot, but there should be SGRs. For example, the events
marked by stars have the same location.
The softest events ($HR < 0.6$ were removed from the GRB sample).}

\end{figure}


 The resulting log N - log P distribution in absolute units is presented in Fig.4

\begin{figure}
  \resizebox{\hsize}{!}{\includegraphics{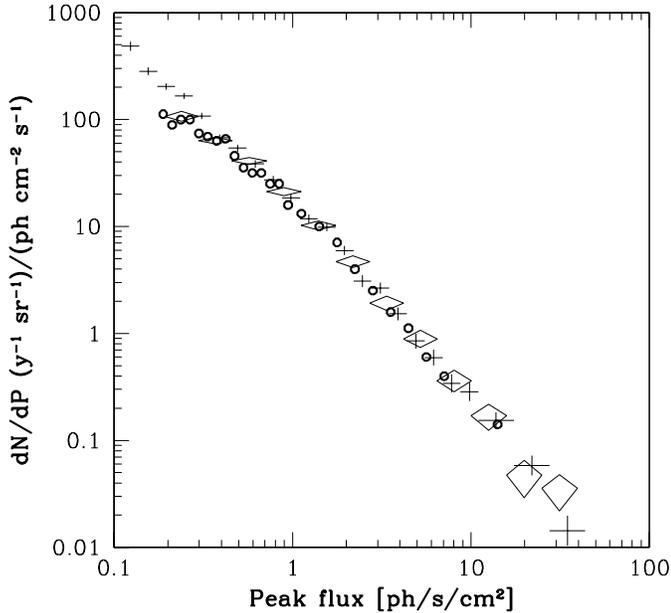}}
\caption{ The differential log N - log P distribution in absolute units for all 
events (2617) (1243 non-triggered) 
detected in the scan.
 Data of Kommers et al.(1998) are shown by circles, BATSE-3 distribution from
... is shown by diamonds (in arbitrary normalisation).
Our data are corrected for the ``test bursts'' efficiency
which is $\sim 0.72$  for bright bursts (due to data gaps and bad 
background intervals) and smoothly declines to the lowest peak fluxes.}
\end{figure}


in comparison with BATSE log N - log P from Meegan et al.  (1998)
(in arbitrary
normalisation) and that from Kommers et al.  (1998) (in absolute units).     
All distributions are normalised with efficiencies estimated by their authors.
Kommers et al.(\cite{KOM98}) have detected events below 0.2 ph s$^{-1}$ cm$^{-2}$, however 
they presented only data with efficiency higher 0.8.  
 The efficiency curve used by Kommers et al.(1998) is a sharper function
of the peak flux. Our log N - log P curve is higher in the range 
$P \sim$ 0.2 - 0.6 ph cm$^{-2}$ s$^{-1}$ because of two factors: 
lower  
efficiency in this range according our estimate and a larger number
 of detected 
events.

\begin{figure}
  \resizebox{\hsize}{!}{\includegraphics{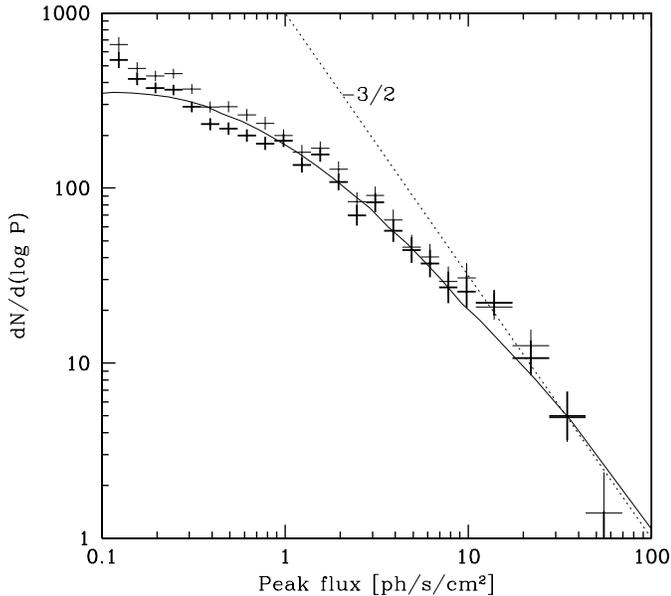}}
\caption{ The differential log N - log P distribution when short bursts are removed 
(thick crosses). (Full distribution,
the same as in the previous Fig., is shown by thin crosses.) 
It cannot be fit with the simplest hypothesis of the 
standard candle non-evolving GRB population: the best fit is shown by 
the solid curve ($\chi^2$ = 50.2 at 22 degrees of freedom). The left most
 data point was
excluded because it could be biased due to the threshold effects.}
\end{figure}


Fig. 5 shows the log N - log P distribution where short events (only one bin 
is above
0.5 of the peak value) were removed from the sample and its best fit with a standard
candle cosmological distribution for non-evolving parent population. 
For the first time the simplest cosmological model cannot fit data.  
 The fit will be even worse 
if we use the star formation rate curve as the GRB evolution scenario.
The rejection of the standard candle hypothesis is not surprising,
however this is still an achievement because the cosmological fit
of the log N --  log P becomes conclusive.


\section {Preliminary conclusions}

 Stern, Poutanen \& Svensson (1997) claimed a possible indication of a turnover
of the log N - log P near BATSE threshold. Such a turnover was also indicated  
by the results of Kommers et al. (1998) . There was a hope to reveal a 
cosmological 
evolution of the GRBs source population, i.e., their likely decline at large z 
together
with the star production rate  (see Totani, 1997  Bagot, Zwart \& 
Yungel'son, 1998).

 The turnover of the log N - log P is not confirmed.
The reason is that the BATSE detection efficiency (see Meegan et al.,1998) 
below 1 ph cm$^{-2}$ s$^{-1}$ turned out to be less than was estimated 
before - the smoothly declining efficiency mimiced a turnover. 
Probably all we see is just a very wide intrinsic luminosity function 
convolved with the cosmological distribution which we will not be able
to extract straightforwardly.
  
Can the range 0.1 - 0.5 ph cm$^{-2}$ s$^{-1}$ be contaminated by 
non-GRB events?

 There is a temptation to suggest that the true GRBs log N - log P is bent 
as the standard candle curve in Fig. 5 and the rise at the left is caused by a 
contamination  with events of another nature. One exciting possibility is a
subpopulation of a relatively nearby bursts: the possible association of 
a supernova and GRB supports this variant. Then log N - log P should bend up
to the Euclidean slope somewhere. Nobody can exlude this,
however weakest events have the same hardness (Fig. 3) as classic GRBs and are 
isotropic in any
frame (Galactic, Solar, Geocentric, where ionospheric events  are 
anisotropic). They have the same range of durations as GRBs and
the same character of the variability.

 A wide intrinsic luminosity function is a more ``economical''
explanation: it must be wide 
and it will fit this log N -- log P easiely. 

\section{Aknowlegements}

 We thank Juri Poutatnen, Aino Skassyrskaia, Andrei Skorbun, Eugeni Stern, 
Vladimir Kurt,
Kirill Semenkov, Stas Masolkin, Alex Sergeev, Max Voronkov, Andrei 
Beloborodov, and Felix Ryde for valuable assistance. 

This work was supported by the Swedish Natural Science Research Council,
the Royal Swedish Academy of Science, the Wennergren Foundation for Scientific
Research, and a NORDITA Nordic Collaboration Project grant. RFBR grant 97-02-16975
supported one of the authors (D.K.).



\end{document}